\documentclass[aps,prb,twocolumn,showpacs,10pt]{revtex4-1}
\usepackage[colorlinks=true,linkcolor=blue,anchorcolor=blue,citecolor=blue]{hyperref}
\usepackage{graphicx}
\usepackage{xcolor}
\usepackage{bm}
\usepackage{scrextend}
\squeezetable

\begin{document}
\title{Effective Coulomb interaction in actinides from linear response approach}

\author{Ruizhi Qiu}
\email{qiuruizhi@itp.ac.cn}
\affiliation{Science and Technology on Surface Physics and Chemistry Laboratory, Mianyang 621908, Sichuan, China}
\author{Bingyun Ao}
\email{aobingyun@caep.cn}
\affiliation{Science and Technology on Surface Physics and Chemistry Laboratory, Mianyang 621908, Sichuan, China}
\author{Li Huang}
\email{lihuang.dmft@gmail.com}
\affiliation{Science and Technology on Surface Physics and Chemistry Laboratory, Mianyang 621908, Sichuan, China}
\date{\today}
\begin{abstract}
The effective on-site Coulomb interaction (Hubbard $U$) between 5$f$ electrons in actinide metals (Th-Cf) is calculated with the framework of density-functional theory (DFT) using linear response approach.
The $U$ values seldom rely on the exchange-correlation functional, spin-orbital coupling, and magnetic states, but depend on the lattice volume and actinide element.
Along the actinide series, the Coulomb parameter $U$ of $\alpha$-phase first decreases slowly, followed by a jump in the vicinity of Pu and then a monotonous increase.
For light actinides, the lattice volume has a sizeable influence on $U$ while the localization of 5$f$ electrons is almost constant.
But for transplutonium metals, $U$ is almost independent of the lattice volume but the electronic localization increases rapidly.
The calculated lattice parameters from DFT+$U$ with the Coulomb parameters as input are in better agreement with the experimental values than those from DFT within local density approximation or Perdew-Burke-Ernzerhof approximation for solids (PBEsol).
In particular, the agreement between PBEsol+$U$ and experiment is remarkable. 
We show that PBEsol+$U$ also well reproduce the experimental bulk moduli and the transition from itinerancy to localization of 5$f$ electrons along the series. 
Therefore it is concluded that DFT+$U$ with $U$ calculated from linear response approach is suitable for a good description of actinide metals.
\end{abstract}
\maketitle

\section{Introduction}
\label{sec:introduction}

Actinide elements produce a variety of fascinating physical behaviors, such as three charge-density wave phases of U metal~\cite{Lander1994}, the unique phonon dispersion of $\alpha$-U~\cite{Manley2001,Manley2006} and $\delta$-Pu~\cite{Lander2003,Dai2003,Wong2003}, unconventional heavy-fermion superconductivity in PuCoGa$_5$~\cite{Sarrao2002}, hidden order phase in URu$_2$Si$_2$~\cite{Mydosh2011}, and high-rank multipolar order in NpO$_2$~\cite{Santini2009}. 
The richness of actinide physics is attributed to the 5$f$ electrons in which the narrow electronic band, complex crystal field splitting, large spin-orbital coupling (SOC), and strong on-site Coulomb interaction met.
A very basic competition of 5$f$ electrons is between itinerancy and localization.
That is to say, 5$f$ electrons have the tendency towards delocalization and participating in the bonding, and the opposite tendency towards localizing around the nuclei and behaving like the partially-filled core electrons. 

\begin{figure}[t]
\begin{center}
\includegraphics[width=0.97\columnwidth]{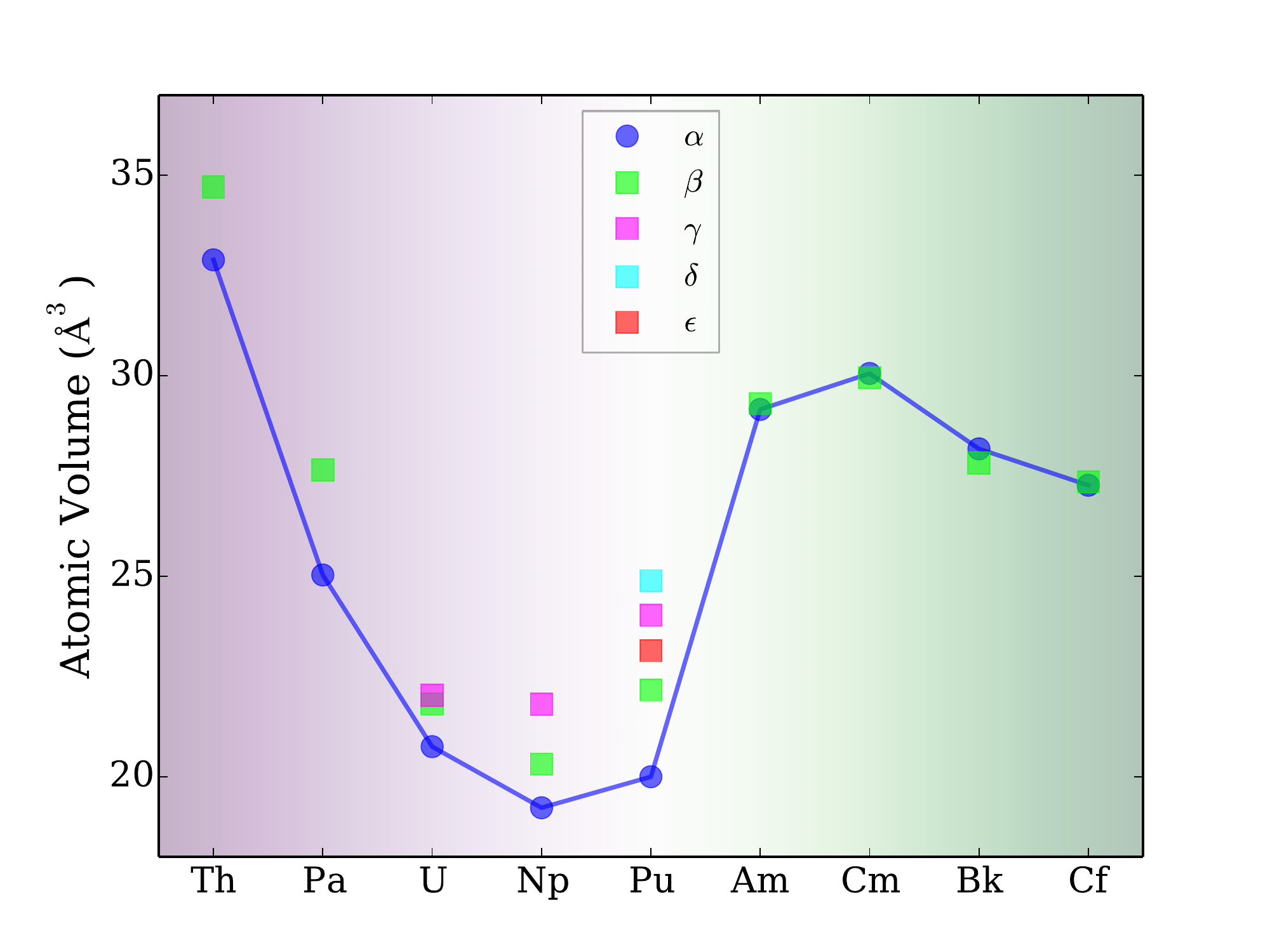}
\caption{Atomic volume of each metal for the actinide series (blue circle) calculated from the crystal structure of $\alpha$-phase. The square block means the crystallographic volume of the other solid allotropic phases of actinides. The data are obtained from Pearson's Handbook~\cite{Villars1997} except $\beta$-Pa~\cite{Marples1965} and $\beta$-Cf~\cite{Heathman2013}.}
\label{fig:volume}
\end{center}
\end{figure}

This competition was highlighted in the atomic volume of the actinides~\cite{Moore2009}, as shown in Fig.~\ref{fig:volume}. 
Along the actinide series, the equilibrium atomic volume first decreases parabolically, then increases abruptly about 43\% from Pu to Am, followed by an almost constant evolution. 
For the actinide metals, the 7$s$ and 6$d$ electronic bands are broad and $s$/$d$ electrons are strongly delocalized, which do not change much along the series. 
The 5$f$ bands are exceedingly narrow, on the order of 2 eV~\cite{Moore2009}.
With the increasing number of $f$ electrons, the parabolic-like behavior of atomic volume for the light actinides (Th-Pu) is indicative of a system with itinerant electrons that are strongly bonding. 
This is similar to that of 5$d$ transition metal series, in which the atomic volume first decreases due to filling of the 5$d$ bonding states and then increases owing to the filling of the antibonding states~\cite{Moore2009}. 
While for the transplutonium elements (Am-Cf), the little volume change with increasing $f$ electrons implies that the 5$f$ electrons are localized. 
This is similar to the 4$f$ rare-earth metals, in which the volume changes little along the 4$f$ series~\cite{Moore2009}. 
Therefore, it is the peculiar 5$f$ states that are the root cause of the anomalous behavior illustrated in Fig.~\ref{fig:volume}.

Apparently, the effective Coulomb interaction is very important for the description of the physical properties of actinide materials.
Now it is well-known that the localization of 5$f$ electrons cannot be reproduced by the basic approximation of exchange-correlation functional, such as local-density approximation (LDA)~\cite{Ceperley1980,Perdew1981} and generalized gradient approximation (GGA)~\cite{Perdew1996,Perdew2008}.
For the actinide compounds with the well-defined localized 5$f$ electrons, such as insulating actinide dioxides (AnO$_2$, An=Pa, U, Np, Pu, Am, and Bk), LDA/GGA predicted the incorrect metallic feature~\cite{Wen2013}.
Various approaches have been proposed to remedy this problem such as self-interaction correction~\cite{Perdew1981},  hybrid functional~\cite{Heyd2003,Heyd2006}, Hubbard correction of intra-atomic Coulomb interaction (DFT+$U$)~\cite{Liechtenstein1995,Dudarev1998}, and dynamical mean-field theory (DFT+DMFT)~\cite{Kotliar2006}.
Due to the clear physical meaning and simple formulation, DFT+$U$ has been extensively used in the first-principles calculation on 5$f$ materials (see Refs.~\onlinecite{Soderlind2010,Wen2013} and references therein).

For most of first-principles DFT+$U$ calculations on actinide materials, this important Coulomb parameter $U$ is usually determined empirically by fitting to existing experimental data.
Actually, the Coulomb parameter could be derived from the spectroscopic measurements by defining $U$ as the energy cost of moving a 5$f$ electron between two atoms both of which initially had $n$ 5$f$ electrons, 2($f^n$) $\rightarrow$ $f^{n-1}$ + $f^{n+1}$. 
Based on the atomic spectral data~\cite{Brewer1971}, the parameters $U$ of actinides were estimated as the energy of the reaction 2($f^nd^2s$) $\rightarrow$ $f^{n-1}d^3s$ + $f^{n+1}ds$~\cite{Herring1967} and 2($f^nds$) $\rightarrow$ $f^{n-1}d^2s$ + $f^{n+1}s$~\cite{Johansson1975} by keeping the charge neutrality.
Note that the atomic limit was used and the metallic situation is the case of real interest.
Later by truncating the free-atom wave function at the Wigner-Seitz radius of the actinide metals and performing the relativistic Hartree-Fock calculations, Herbst {\it et al.}~\cite{Herbst1976} corrected the values of $U$, i.e., the energy of the reaction 2($f^nd^{m-1}s$) $\rightarrow$ $f^{n-1}d^{m}s$ + $f^{n+1}d^{m-2}s$, with the energy difference between free-atom and Wigner-Seitz cell calculation. 
All these estimated values are listed in Table~\ref{tab:U} and there is a general increase in $U$ from Th to Am, which implies that the effective Coulomb interaction is increased and the hopping is progressively impeded across the actinide series. 

Nowadays it is possible to compute the Hubbard $U$ from first-principles and the methods include constrained random-phase approximation (cRPA)~\cite{Aryasetiawan2004,Aryasetiawan2006}, constrained DFT~\cite{Gunnarsson1989,Anisimov1991}, and linear response approach~\cite{Cococcioni2005}.
In the cRPA, the Hubbard $U$ is obtained as the expectation value of Coulomb interaction on the wave functions of the localized basis set.
As this approach allows for the calculations of the matrix elements of the Hubbard $U$ and its energy dependence, it becomes particularly popular within the DFT+DMFT community.
Recently Amadon~\cite{Amadon2016,Amadon2018} use the cRPA to estimate the Hubbard $U$ of actinide metals, which are also listed in Table~\ref{tab:U}.
For Pu and Am, the Hubbard $U$'s are much smaller than the typical width of $f$ bands (2.0 eV), which means that they are systems with weak electronic correlations. 
The Hubbard $U$'s of Pu and Am are much smaller than the typical width of f bands, 2.0 eV~\cite{Moore2009}.
This indicates that the electronic correlations of Pu and Am are weak, being in contradiction to the photoemission experiments~\cite{Gouder2001,Naegele1984}.
In addition, the $U$ values from cRPA are adaptable to the DFT+DMFT calculation but less transferable to DFT+$U$ due to the use of Wannier functions in the cRPA calculation.

For constrained DFT, the Hubbard $U$ is obtained from the total-energy variation with respect to the occupation number of the localized orbitals, which is identified as the shift in the Kohn-Sham eigenvalues by virtue of Janak theorem~\cite{Janak1978}.
The implementation of DFT using the localized basis sets makes it possible to change the occupation of the localized orbitals but fails to perform the screening of other delocalized orbitals.
A further improvement of constrained DFT was the linear response approach, in which one can utilize the pseudopotential methods using plane wave basis sets and not just the localized basis sets.
The localized orbitals are perturbed by a single-particle potential and the Hubbard $U$ is determined by using the density response functions of the system with respect to these localized perturbations.
The scheme is internally self-consistent and widely used in the first-principles calculation community~\cite{Himmetoglu2014}.
To our best knowledge, computation of $U$ using linear response approach has not been performed to pure actinide except uranium~\cite{Xie2013}.
In the work we apply the linear response approach to systematically calculate the Hubbard $U$ of actinide metals. 
This trend of $U$ along the actinide series will improve our understanding of the 5$f$ electron behavior. 
In addition, DFT+$U$ with $U$ from linear response approach could be assessed for the description of the physical properties of actinide metals.
The rest of the paper is organized as follows.
The computational models, i.e., various phases of the actinide metals, are introduced in the appdenix~\ref{sec:model}.
Computational method and details are given in Sec.~\ref{sec:method}.
Section~\ref{sec:interaction} presents the results of Coulomb parameters $U$ and discussion.
The assessment of DFT+$U$ with Coulomb parameters as input is shown in Sec.~\ref{sec:property}.
The last section summarizes the main achievements of this work.


\begin{table}[t]
\caption{List of Coulomb parameters $U$ of actinides in the literature.}
\label{tab:U}
\begin{tabular}{lllllllll}
\hline
Metal & Th & Pa & U & Np & Pu & Am & Cm & Bk \\
\hline
Ref.~\onlinecite{Herring1967} & 1.7 & 2.0 & 2.6 & 2.7 & 3.2 & 3.0 & - & - \\
Ref.~\onlinecite{Johansson1975} & 1.5 & 1.6 & 2.3 & 2.6 & 3.5($\pm$1) & 5($\pm$1) & - & - \\
Ref.~\onlinecite{Herbst1976} & 3.0 & 3.5 & 3.7 & 3.7 & 4.0 & 4.2 & 9.5 & 3.4 \\
Ref.~\onlinecite{Lukoyanov2014} & 2.3 & - & - & - & - & - & - & -  \\
Ref.~\onlinecite{Xie2013} & - & - & 1.87-2.1 & - & - & - & - & - \\
Ref.~\onlinecite{Amadon2016} & - & - & 0.8 & 1.0 & 0.95 & 1.5 & 3.4 & - \\
\hline
\end{tabular}
\end{table}

\section{Computational method}
\label{sec:method}

All calculations have been performed using the Vienna Ab-initio Simulation Package (VASP)~\cite{Kresse1996}.
The ion-electron interaction was described using the projector augmented wave (PAW) formalism~\cite{Blochl1994,Kresse1999}, which has the accuracy of all electron methods because it defines an explicit transformation between the all-electron and pseudopotential wave functions by means of additional partial-wave basis functions~\cite{Lejaeghere2016}.

\subsection{DFT+$U$ formalism}

DFT+$U$ is based on a corrective functional inspired to the Hubbard model and thus is one of the simplest approach to improve the description of the ground state of correlated system.
The variable of the corrective functional is the occupation number matrix of the localized orbitals (${\bm n}^{I}$), which is computed from the projection of Kohn-Sham orbitals ($\psi^\sigma_{{\bm k}\nu}$) into the states of localized basis set of choice ($\phi^I_m$, e.g., atomic orbitals in VASP):
\begin{eqnarray}
    n^{I\sigma}_{mm'}=\sum_{{\bm k}\nu}f^\sigma_{{\bm k}\nu} \left\langle\phi^I_m\vert\psi^\sigma_{{\bm k}\nu}\right\rangle \left\langle\psi^\sigma_{{\bm k}\nu}\vert\phi^I_{m'}\right\rangle.\label{eq:occupation}
\end{eqnarray}
Here $\sigma$, ${\bm k}$, $\nu$, $I$, and $m$ are spin, Bloch wavevector, band index, Hubbard atom index, and index of localized orbitals, respectively; $f^\sigma_{{\bm k}\nu}$ is the Fermi-Dirac distribution of the Kohn-Sham states. 
Note that the trace of occupation number matrix, $n^I={\rm Tr}[{\bm n}^I] = \sum_{\sigma m}n^\sigma_{mm}$, is the total occupation number of localized orbital in Hubbard atom $I$.
If the localized orbitals are set as the $f$ atomic orbitals, the index $m$ might be understood as the magnetic quantum number, $m=-3,-2,\ldots,3$.

Here the DFT+$U$ corrective energy functional is chosen as the simplified rotationally invariant form~\cite{Dudarev1998,Cococcioni2005}, i.e.,
\begin{eqnarray}
    E_U[\{{\bm n}^{I}\}] = 
    \sum_I\frac{U^I}{2}{\rm Tr}\left[{\bm n}^I\left(1-{\bm n}^I\right)\right].\label{eq:hubbard}
\end{eqnarray}
Note that the trace also includes the summation of spin $\sigma$.
This simplified version is equivalent to the fully localized limit of fully rotationally invariant form~\cite{Liechtenstein1995} with the exchange parameter being zero.
It has been successfully applied to many materials and yields similar results as the fully one~\cite{Himmetoglu2014}, such as plutonium dioxide~\cite{Sun2008,Jomard2008}.




\subsection{Linear response approach}

As expected, the results from DFT+$U$ sensitively depend on the value of Hubbard $U$ from the expression of the corrective functional (\ref{eq:hubbard}).
From a conceptual point of view, it is not satisfactory to tune $U$ to seek agreement with available experimental results.
More importantly, semi-empirical choice of $U$ is unable to consider the variations of Hubbard $U$ during chemical reactions and structural/magnetic transition~\cite{Himmetoglu2014}.
Thus it is important to compute the Hubbard $U$ in a consistent and reliable way.
Note that $U$ should be calculated for every Hubbard atom for the considered crystal structure and the specific magnetic ordering of interest.

The linear response approach aimed at computing the Hubbard $U^I$ as the second derivative of the ground state total energy with respect to the occupation number $n^I$.
First of all, the single-body potential is perturbed by an external potential that only acts on the localized orbitals of a Hubbard atom $I$, 
$\alpha^I\vert\phi^I_m\rangle\langle\phi^I_m\vert$,
in which $\alpha^I$ is the amplitude of the perturbation on Hubbard atom $I$.
Solving the modified Kohn-Sham equations with the perturbed single-body potential yields an $\alpha^I$-dependent ground state total energy
\begin{eqnarray}
    {\cal E}[\{\alpha^I\}]=\min_{{\bm \rho}({\bm r})}\left\{E_{\rm DFT}[{\bm \rho}({\bm r})]+\alpha^In^I\right\},
\end{eqnarray}
with ${\bm \rho}({\bm r})$ being the electron density and $E_{\rm DFT}[\cdot]$ being the total energy function of pure DFT.
Then Legendre-Fenchel transformation is used to transform this optimization problem into the corresponding dual problem.
That is to say, an $n^I$-dependent ground state total energy could be recovered,
\begin{eqnarray}
    E[\{n^I\}] & = & {\cal E}[\{\alpha^I\}]-\alpha^I \left(\frac{\partial {\cal E}[\{\alpha^I\}]}{\partial \alpha^I}\right) \nonumber\\
    & = & {\cal E}[\{\alpha^I\}]-\alpha^I n^I.
\end{eqnarray}
Based on this definition, the first derivative of the total energy with respect to $n^I$ is given by
\begin{eqnarray}
    \frac{\partial E[\{n^I\}]}{\partial n^I} = -\alpha^I,
\end{eqnarray}
and the second derivative,
\begin{eqnarray}
    \frac{\partial^2 E[\{n^I\}]}{\partial (n^I)^2} = -\frac{\partial \alpha^I}{\partial n^I}.
\end{eqnarray}
In actual numerical calculation, all $\{n^J\}$ vary in response to the change of $\alpha^I$ and the evaluated quality should be the response function $\chi$ whose matrix element is evaluated from the finite differences,
\begin{eqnarray}
    \chi_{IJ}=\frac{\delta n^I}{\delta \alpha^J}.
\end{eqnarray}
In addition, the electronic wave function of non-interacting electron systems would be reorganized in response to the perturbation and this response should be eliminated since it is not related to the electron-electron interaction. 
This response function $\chi_0$ could be evaluated by performing a non-self-consistent electronic structure calculation with the charge density being kept constant and collecting the response of this non-interacting system in terms of variation of all $\{n^J\}$.
Finally the Hubbard $U^I$ is given by 
\begin{eqnarray}
    U^I=(\chi_0^{-1}-\chi^{-1})_{II}.
\end{eqnarray}

\subsection{Computational parameters}

We used the official PAW pseudopotentials for Th, Pa, U, Np, Pu, Am, Cm, and Cf, which were generated with reference valence configurations, 6$s^2$6$p^6$7$s^2$5$f^1$6$d^1$, 6$s^2$6$p^6$7$s^2$5$f^2$6$d^1$, 6$s^2$6$p^6$7$s^2$5$f^3$6$d^1$, 6$s^2$6$p^6$7$s^2$5$f^4$6$d^1$, 6$s^2$6$p^6$7$s^2$5$f^5$6$d^1$, 6$s^2$6$p^6$7$s^2$5$f^6$6$d^1$, 6$s^2$6$p^6$7$s^2$5$f^7$6$d^1$, and 6$s^2$6$p^6$7$s^2$5$f^8$6$d^2$, respectively. 
The pseudopotential of Bk is not constructed and thus Bk is not included in our calculation.
The energy levels of 5$f$ and 6$d$ are very close to each other and their occupations will be redistributed in the actual numerical calculation.
In this study, we have compared the following exchange-correlation functionals: LDA~\cite{Ceperley1980,Perdew1981}, GGA of Perdew-Burke-Ernzerhof (PBE)~\cite{Perdew1996}, and PBE revised for solids (PBEsol)~\cite{Perdew2008}.
The partial occupancies for each Kohn-Sham wave function are determined by the tetrahedron method with Bl\"ohl correction.
For the convergence tests, the cutoff energy for plane wave basis and the Monkhorst-Pack (M-P)~\cite{Monkhorst1976} $k$ point meshes are determined.
In this work, all the cutoff energy is chosen as 450 eV except Cf, which is set as 600 eV.
A 9$\times$9$\times$9 M-P $k$-point meshes are used for fcc, bcc, and bct phases.
For dhcp-phase, $\alpha$-U, $\beta$-U, $\alpha$-Np, $\beta$-Np, $\alpha$-Pu, $\beta$-Pu, and $\gamma$-Pu, the 9$\times$9$\times$3, 11$\times$7$\times$7, 3$\times$3$\times$7, 7$\times$7$\times$7, 7$\times$7$\times$9, 5$\times$7$\times$3, 3$\times$3$\times$3, and 9$\times$5$\times$3 M-P $k$ point meshes are chosen.
SOC is considered in some calculations of Hubbard $U$ for the discussion of the effect of SOC on the effective Coulomb interaction.
For the physical properties, all the calculations are performed by including SOC.
The magnetic orders considered here include diamagnetic (DM), ferromagnetic (FM) and non-collinear antiferromagnetic (AFM) and will be considered case by case.

For the calculation of Hubbard $U$, the response function matrix $\chi$ and $\chi_0$ are derived by setting $\alpha$ parameter to $\pm0.1$, $\pm0.2$, $\pm0.3$, $\pm0.4$, and $\pm0.5$. The parameters are so chosen to remove the numerical noise. 
Only $N_{\rm Wyckoff}$ column of $\chi$ and $\chi_0$ is extracted from the calculation with $N_{\rm Wyckoff}$ being the number of Wyckoff positions in the cell. 
All the other matrix elements are reconstructed by symmetry.
The presented $U$ is chosen as the $U^I$ of the first Hubbard atom.
Considering the periodicity of DFT calculation in solid, the local perturbation should not be overlapped and thus a supercell approach is adopted here.
For fcc, bcc, bct phase, $\alpha$-U, $\beta$-Np, we used a 2$\times$2$\times$2 supercell. 
For dhcp phase and $\gamma$-Pu, a 2$\times$2$\times$1 supercell is used.
A 1$\times$2$\times$2 supercell is constructed for $\alpha$-Np and the unit cell is used for $\beta$-U, $\alpha$-Pu and $\beta$-Pu.
The calculated $U$ using the above supercell and a larger supercell are within about 0.05 eV.

\section{Effective Coulomb interaction}
\label{sec:interaction}

\begin{table}
\begin{center}
\caption{Calculated Hubbard $U$ (in units of eV) of actinides from different computational schemes.}
\label{tab:Hubbard}
\begin{tabular}{lllllll}
\hline
& Structure & Magnetic & LDA & PBE & PBEsol & LDA+SOC \\
\hline
$\alpha$-Th   & fcc        & DM  & 2.6363 & 2.4619 & 2.4677 & 2.6318 \\ 
$\beta$-Th    & bcc        & DM  & 2.6324 & 2.5254 & 2.5160 &  \\
$\alpha$-Pa   & bct        & DM  & 2.5553 & 2.5433 & 2.5239 & 2.6008 \\ 
$\beta$-Pa    & bcc        & DM  & 2.6675 & 2.6364 & 2.6326 &  \\
	      & fcc        & DM  & 3.0344 & 2.9337 & 2.9500 &  \\
$\alpha$-U    & $Cmcm$     & DM  & 2.3954 & 2.3928 & 2.3913 & 2.4082 \\
$\beta$-U     & $P4_2/mnm$ & DM~\footnote{\label{fn:mag}The magnetic moment per atom is about 0$\sim$0.3 $\mu_B$, being small but not negligible in the numerical calculation.}   & 2.9071 & 2.8578 & 2.8829 &  \\
$\gamma$-U    & bcc        & DM  & 2.5301 & 2.4792 & 2.5264 &  \\
$\alpha$-Np   & $Pnma$     & DM  & 2.2026 & 2.2146 & 2.2145 &  \\
$\beta$-Np    & $P42_12$   & DM  & 2.3972 & 2.4291 & 2.4104 &  \\
$\gamma$-Np   & bcc        & AFM~\footnote{The different magnetic states are sorted by the ascending order of total energy.} 
				 & 2.4583 & 2.4538 & 2.4446 &  \\
              &            & FM  & 2.5081 & 2.4782 & 2.4975 &  \\
              &            & DM  & 2.4091 & 2.4369 & 2.4845 &  \\
$\alpha$-Pu   & $P2_1/m$   & DM~\footref{fn:mag}
				 & 2.0675 & 2.0263 & 2.0898 & \\
$\beta$-Pu    & $C2/m$     & FM  & 2.4635 & 2.4786 & 2.4740 & \\
$\gamma$-Pu   & $P6_3/m$   & DM  & 2.7069 & 2.5982 & 2.6923 & \\
$\delta$-Pu   & fcc        & AFM & 2.7584 & 2.8015 & 2.8086 & 2.7307 \\
              &            & FM  & 2.7008 & 2.6732 & 2.7466 &  \\
              &            & DM  & 2.7792 & 2.6917 & 2.6059 &  \\
$\epsilon$-Pu & bcc        & DM  & 2.5551 & 2.5582 & 2.5897 & \\
$\alpha$-Am   & dhcp       & AFM & 3.2112 & 3.2064 & 3.2146 & \\
              &            & FM  & 3.2771 & 3.2623 & 3.2491 &  \\
$\beta$-Am    & fcc        & AFM & 3.2969 & 3.2416 & 3.2454 &  \\
              &            & FM  & 3.4172 & 3.3333 & 3.3256 &  \\
$\alpha$-Cm   & dhcp       & AFM & 3.6648 & 3.4314 & 3.4699 &  \\
              &            & FM  & 3.2992 & 3.2698 & 3.2857 &  \\
$\beta$-Cm    & fcc        & AFM & 3.5959 & 3.2582 & 3.3144 & \\
              &            & FM  & 3.2708 & 3.2218 & 3.3530 & \\
$\alpha$-Cf   & dhcp       & FM  &        & 4.3567 & 4.5242 & \\
              &            & AFM &        & 4.8949 & 4.1342 & \\
$\beta$-Cf    & fcc        & FM  &        & 4.2423 & 4.2732 & \\
              &            & AFM &        & 4.4180 & 4.2758 & \\ 
\hline
\end{tabular}
\end{center}
\end{table}

\subsection{Influence of computational formulation}

For actinide metals at different phases, the calculated values of effective Coulomb interaction $U$ within different formulation are listed in Table~\ref{tab:Hubbard}.
Before attempting to examine trends in $U$ along the series, it is instructive to consider the influence of exchange-correlation functional, SOC and magnetic order.

First, the difference in $U$ between PBE and PBEsol is negligible for most phases.
Statistical analysis reveals that the deviation is smaller than 0.02 eV for half phases considered. 
The largest difference lies in Cf, whose pseudopotential was constructed recently and required to be carefully inspected.
The reason of negligible difference is that PBEsol is revised to improve the equilibrium properties of solids but holds the good description of free atom~\cite{Perdew2008}.
Analysis also shows that in general, $U$ from PBEsol is slightly larger than that from PBE.
The difference in $U$ between LDA and PBE is small but not negligible.
Statistical analysis reveals that the deviation is smaller than 0.05 eV for more than half phases considered. 
In general, $U$ from LDA is slightly larger than that from PBE, which may owing to the overbinding usually found in LDA.
Overall, the effect of exchange-correlation functional on the value of $U$ is small for actinide metals.
Even for the most complex low-symmetry metals, $\beta$-Pu, $\alpha$-Pu, and $\beta$-U, the results from LDA/PBE/PBEsol are so close.
This consistent behavior of $U$ illustrates the reliability of our calculation.
\begin{figure}[t]
\begin{center}
\includegraphics[width=0.97\columnwidth]{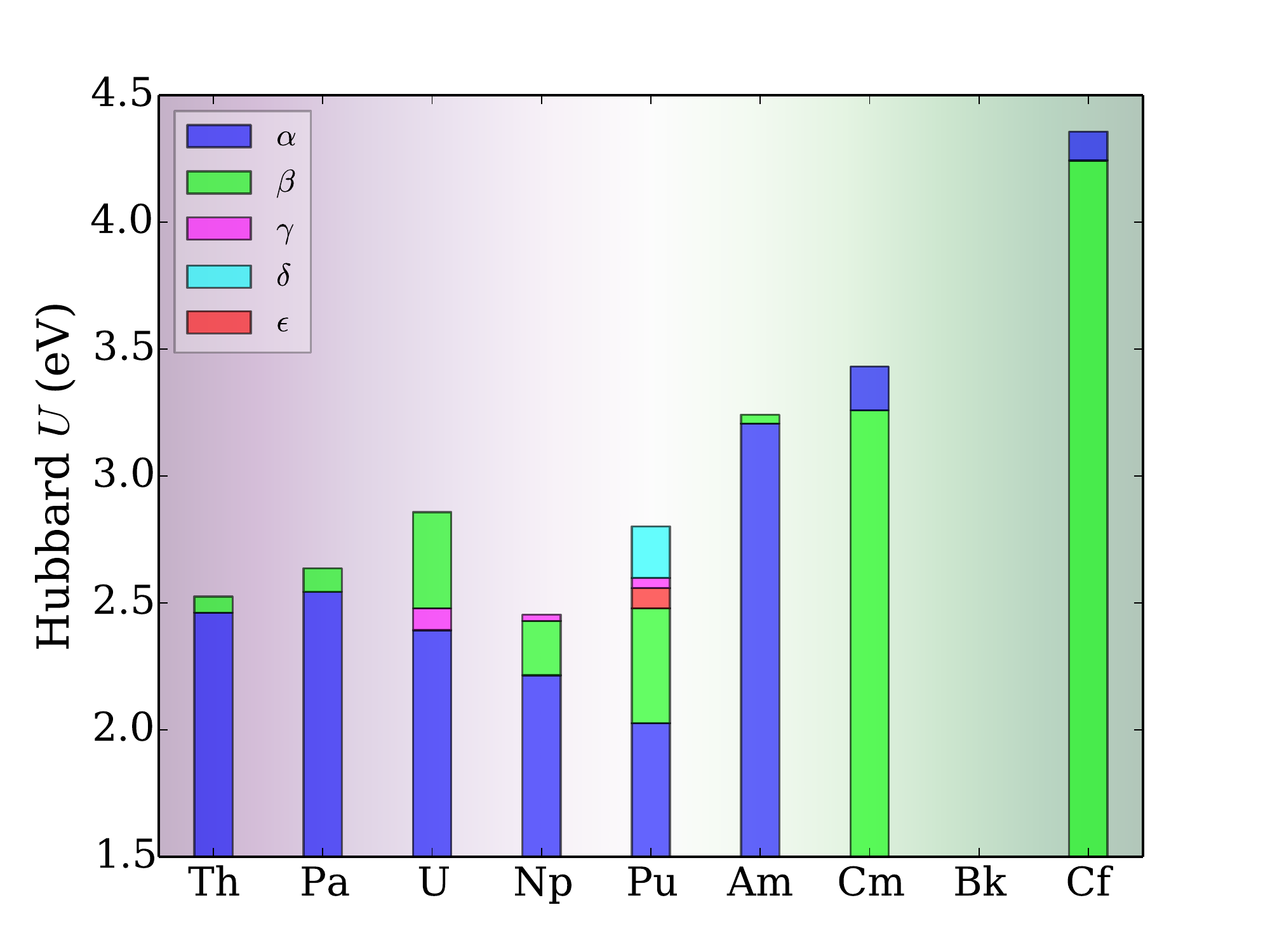}
\caption{Calculated effective Coulomb interaction of PBE+$U$ for the actinide metal series. For Cm and Cf, the estimated $U$ of $\beta$-phase is less than that of $\alpha$-phase, resulting the different main color.}
\label{fig:hubbard}
\end{center}
\end{figure}

Second, the effect of SOC is also negligible even for low-symmetry system $\alpha$-U.
This could be understood by noting that the Hamiltonian of SOC is a single body term to employ the scalar relativistic correction. 
In the presence of SOC, the angular and spin part of single-electron wave function for a free atom will be greatly altered but the radial part, which mainly contributes to the effective on-site Coulomb interaction, was unchanged.
Thus we conclude that the effect of SOC on $U$ is very small for actinides.
In the following non-collinear calculations of physical properties, the input $U$'s are thus chosen as that from spin-polarized calculation.

Third, most of the lowest-energy magnetic states are DM and thus the comparison of different magnetic states could not be made.
For the magnetic systems from first-principle calculation, the calculated $U$ from AFM, FM, and DM states are compared.
The difference of $U$ among AFM, FM, and DM is smaller than 0.2 eV for most allotropes.
The large differences lie in the transplutonium metals, whose 5$f$ electrons are localized and give rise to the magnetic moment spontaneously.
As a whole, the influence of computational formulation is limited.

\subsection{Trend of Hubbard $U$ along the series}

Now let us turn to the investigation of the trend in $U$ along the actinide series.
To illustrate this point, we plot the values of $U$ from PBE with respect to the actinde elements in Fig.~\ref{fig:hubbard}.
First of all, let us focus on the $\alpha$-phase of actinide metals.
For the $\alpha$-phase, the calculated $U$ decrease slowly for light actinides (Th-Pu) while for transplutonium element (Am-Cf), $U$ increases rapidly.
There is a jump in $U$ occuring between Pu and Am, which is very similar to the jump of the atomic volume in Fig.~\ref{fig:volume}.
The 50\% increase of the effective Coulomb interaction from $\alpha$-Pu to $\alpha$-Am is an indication of electron localization.
Since the on-site Coulomb interaction $U$ could be defined as the one-center Coulomb integral between the localized wave functions in the Hubbard model, the degree of localization implies the increase of the integral and also the increase of $U$.
But for lighter actinides, the decrease of $U$ does not implies the decrease of the degree of 5$f$ localization.
The dominant factor is the parabolical decrease of atomic volume, as shown in Fig.~\ref{fig:volume}.
To clarify this, we performed first-principles calculation of $U$ on the fcc An (An=Th-Cm,Cf) with the same lattice parameter 5.0~\AA~using PBE.
It was found that $U$ has an almost constant value from Pa to Pu. 
The jump from Pu to Am also appears in this model calculation.
For Th, the effective Coulomb interaction is smaller than other actinide metals due to its small 5$f$ occupation.
For transplutonium elements, the effective Coulomb interaction, which is no longer influenced by the atomic volume, increases monotonously from Am to Cf.
This increase is contributed from the increase of the degree of 5$f$ localization, which is partly manifested in the magnetic properties of transplutonium metals~\cite{Smith1982}.

\begin{figure}[t]
\begin{center}
\includegraphics[width=0.97\columnwidth]{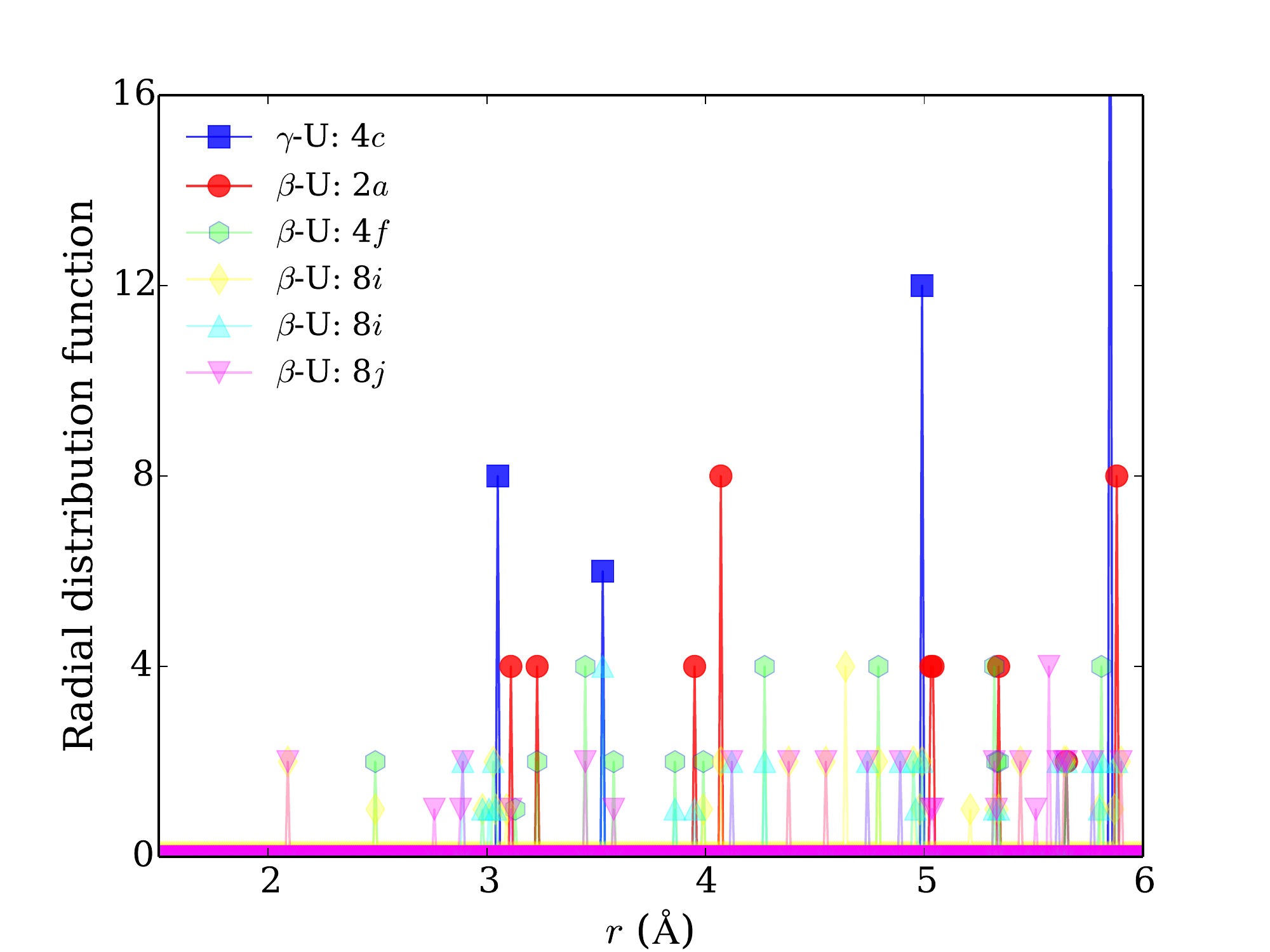}
\caption{The radial distribution function of $\gamma$-U and $\beta$-U with respect to different Wyckoff positions.}
\label{fig:Urdf}
\end{center}
\end{figure}

Next let us focus on the high-temperature phases of actinide metals.
For the light actinides, the atomic volume of high-temperature phases is larger than that of room-temperature $\alpha$-phase due to the thermal expansion.
Accordingly, the calculated $U$ of high-temperature phases is larger than that of $\alpha$-phase.
The larger the atomic volume is, the larger the effective Coulomb interaction is.
Specifically for the five allotropic phases of Pu, the order of $U$ is same as that of atomic volume, as shown in Fig.~\ref{fig:volume} and \ref{fig:hubbard}.
This is consistent with the valence-band photoemission spectra for $\alpha$-Pu and $\delta$-Pu~\cite{Gouder2001}.
Furthermore, the variation range of $U$ widen from Th to Pu, which also results from the variation of the atomic volume.

The only surprise comes from the large value of $U$ for $\beta$-U.
To understand this, we plot the radial distribution function (RDF) of $\beta$-U for five atoms at different Wyckoff positions in Fig.~\ref{fig:Urdf}.
The RDF of $\gamma$-U is also presented for comparison.
The first five neighbors relative to the atom at 2$a$ position are far from those of $\gamma$-U.
Since 2$a$ position is an important symmetry site, it may dominate the U-U distance and thus results in the increase of effective Coulomb interaction.

For transplutonium elements, the effect of allotropy on $U$ is not significant since the lattice volume varies within in a small range (see Fig.~\ref{fig:volume}).
For Cm and Cf, the calculated $U$ of $\beta$-phase is even smaller than that of $\alpha$-phases.
We also calculate $U$ of bcc An with the same volume as fcc An to consider the effect of crystal structure on $U$ but doesn't find a general rule.


\subsection{Comparison with the previous results}

Finally let us compare with the Coulomb parameters of actinides in the literature, as shown in Table~\ref{tab:U}.
For Th-Am, our results are close to but smaller than the estimation by Herbst {\it et al.}~\cite{Herbst1976}.
Since the Hartree-Fock method overestimate spin/orbital polarization, a small $U$ should be used to compensate this effect.
The difference suggests that it may be not optimal to use theoretical $U$ directly in DFT+$U$ calculations.
Of course, the empirical $U$ is still useful and can provide guidelines for the theoretical $U$.

For fcc Th, the value $U$ in this work is close to that in Ref.~\onlinecite{Lukoyanov2014} which is calculated using constrained DFT~\cite{Gunnarsson1989}.
The small difference may result from the absence of the screening of other delocalized $s$ and $d$ orbitals in the constrained DFT.
For U, our PBE results are consistent to that in Ref.~\onlinecite{Xie2013} which also use the linear response approach.
The discrepancy is due to the different computational parameters.
For $\beta$-U, the effect of different Wyckoff positions on the calculation of $U$ may be not considered in Ref.~\onlinecite{Xie2013}.

Our results differ strongly from that from cRPA calculation~\cite{Amadon2016}, whose Coulomb parameters are too small.
We suspect that the metallic feature of actinide influenced the localization of the constructed Wannier orbitals and then underestimated the on-site Coulomb repulsion.
The results of cRPA calculations are very sensitive to the choices of the outer energy window in which the correlated orbitals are defined, and the inner energy window in which the electronic transitions are suppressed for the calculations of constrained polarization~\cite{Aryasetiawan2004,Aryasetiawan2006}.

Considering the order of the typical bandwidth of $f$ bands in actinide metals, 2 eV~\cite{Moore2009},
our calculated values of $U$ for light actinides lie between 2.0-3.0 eV, which implies that the light actinides are system with intermediate electronic correlation. 
For $\delta$-Pu and transplutonium metals, the electronic correlation is strong.
Since LDA/GGA often fails to describe systems with intermediate and strong electronic correlation, the effective Coulomb interaction should be consider in the first-principle calculations, which is the content of next section.

\begin{figure}[t]
\begin{center}
\includegraphics[width=0.97\columnwidth]{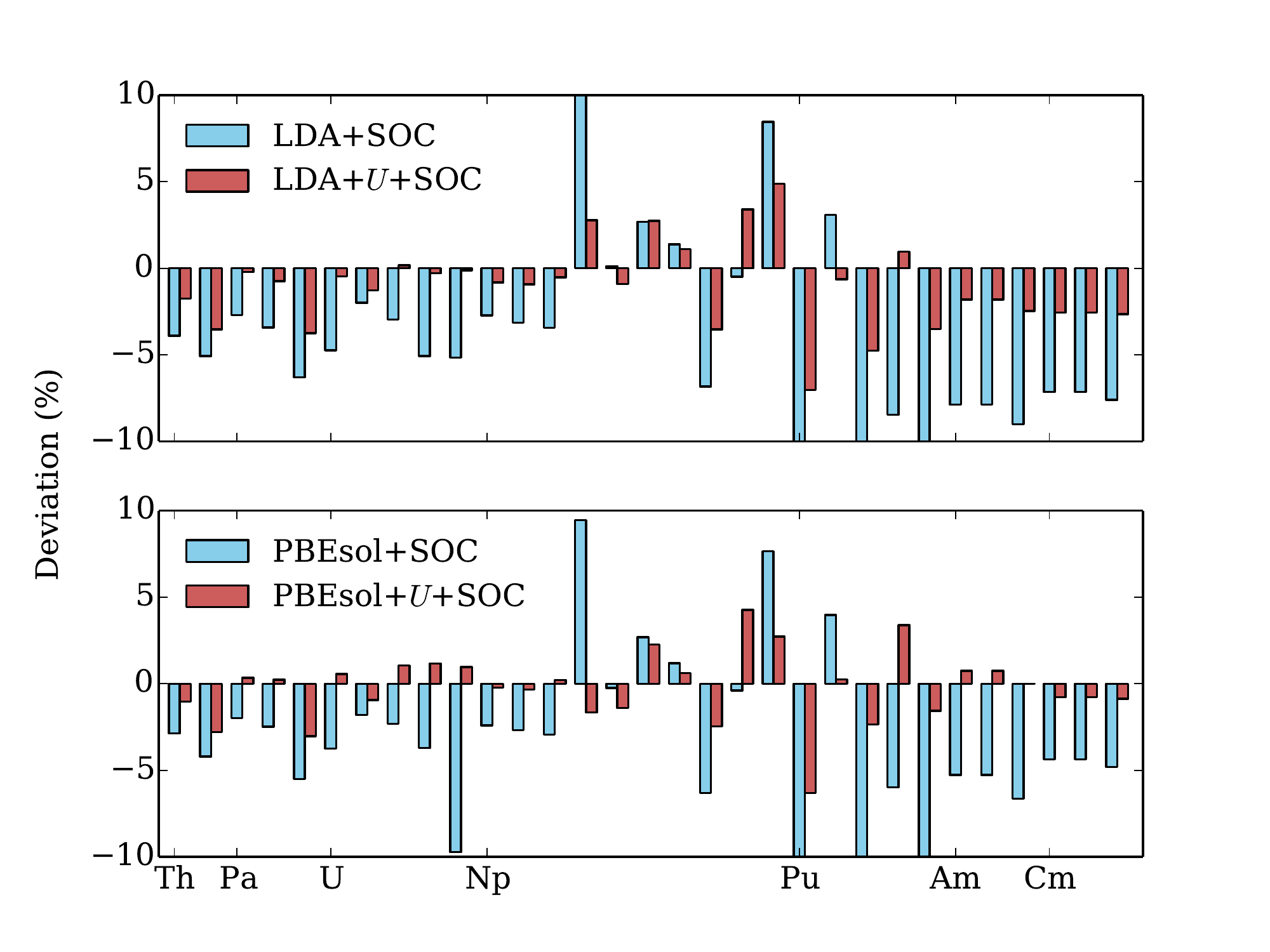}
\caption{The deviation of lattice parameters (from left to right: $\alpha$-Th, $\beta$-Th, $\alpha$-Pa, $\beta$-Pa, $\alpha$-Np, $\beta$-Np, $\gamma$-Pu, $\delta$-Pu, $\epsilon$-Pu, $\alpha$-Am, $\beta$-Am, $\alpha$-Cm, $\beta$-Cm) calculated using LDA/PBEsol+SOC and LDA/PBEsol+$U$+SOC from the experimental values. For specific lattice parameters, please see the appendix~\ref{sec:model}.}
\label{fig:lattice1}
\end{center}
\end{figure}

\section{Physical properties}
\label{sec:property}

\begin{figure}[t]
\begin{center}
\includegraphics[width=0.97\columnwidth]{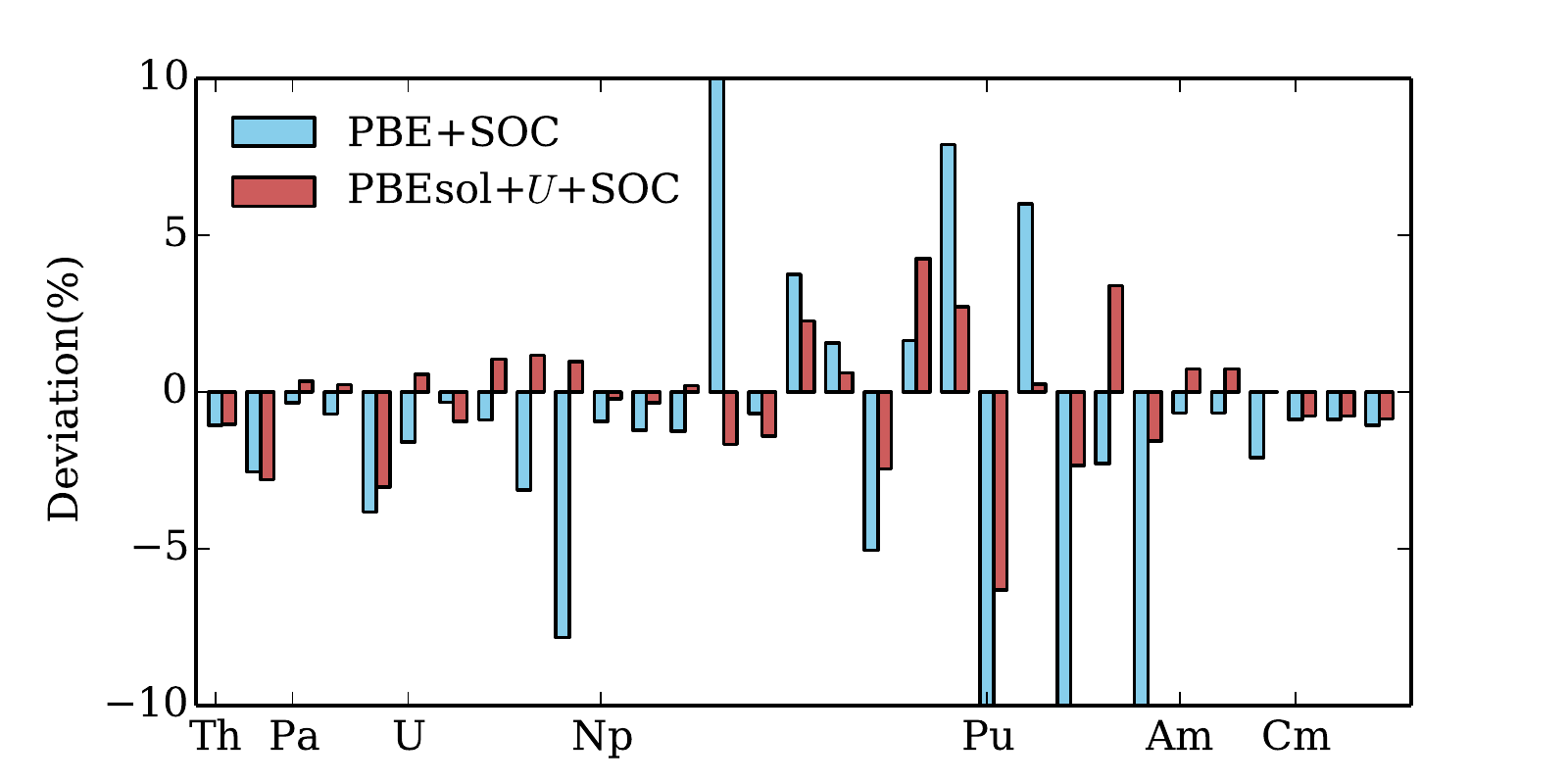}
\caption{The deviation of lattice parameters (from left to right: $\alpha$-Th, $\beta$-Th, $\alpha$-Pa, $\beta$-Pa, $\alpha$-Np, $\beta$-Np, $\gamma$-Pu, $\delta$-Pu, $\epsilon$-Pu, $\alpha$-Am, $\beta$-Am, $\alpha$-Cm, $\beta$-Cm) calculated using PBE+SOC and PBEsol+$U$+SOC from the experimental values. For specific lattice parameters, please see the appendix~\ref{sec:model}.}
\label{fig:lattice2}
\end{center}
\end{figure}

After calculating the Coulomb parameter $U$ using linear response approach, it is quite natural to check if DFT+$U$ could provide a better description of the physical properties of actinide metals than pure DFT.
First, let us compare the theoretical lattice parameters with the experimental values.
To obtain the optimal lattice parameters of actinide metals, we first perform the structural relaxation using the conjugate-gradient and quasi-Newton algorithm.
As mentioned above, SOC are considered in all calculations and the relaxations keep the experimental structural symmetry and the original magnetic state.
Within the framework of DFT+$U$, the lattice should be relaxed using structurally consistent Hubbard $U$~\cite{Hsu2009} since $U$ is dependent on the lattice spacing.
However, the difference of $U$ between the optimized and experimental structure is very small and then the deviation of lattice parameters using constant $U$ from that using structurally consistent $U$ is negligible.
Thus for simplification, the computational scheme is chosen as DFT+$U$ with constant $U$ from Table~\ref{tab:Hubbard}.
The presented optimal lattice parameters and bulk moduli are obtained by fitting the energy-volume data with the third-order Birch-Murnaghan equation of state (EOS)~\cite{Birch1947}.
This procedure has to keep the continuity of the energy-volume curve, which partially rules out the presence of metastable states in the magnetic and DFT+$U$ calculation.
Note that the internal parameters of low-symmetry structures are determined only by the structural relaxation.
For $\beta$-U, $\alpha$-Pu and $\beta$-Pu, it is difficult to obtain the fully-relaxed structures due to the presence of too many internal lattice parameters.
Thus their results are not present in the following.

The calculated lattice parameters are presented in terms of the deviation from the experimental results.
Some experimental results are evaluated under high temperature but still could provide certain reference significance.
The deviations within LDA/PBEsol and LDA/PBEsol+$U$ are plotted in Fig.~\ref{fig:lattice1}.
For LDA, the lattice parameters are often underestimated due to the overbinding. 
As pointed in Ref.~\onlinecite{Soderlind2010b} and confirmed here, the performance of PBEsol is similar to LDA for actinide metals.
For systems with slowly varying electronic densities, LDA and PBEsol are a good choice but not for the actinide metals in which the electronic density varies rapidly.
PBE could remedy this discrepancy and here it is found that Hubbard correction has the same feature.
As can be seen from Fig.~\ref{fig:lattice1}, the deviations are greatly suppressed by the inclusion of the effective Coulomb interaction.

Since both PBE and LDA/PBEsol+$U$ could suppress the deviations of LDA/PBEsol, it is necessary to compare the performance of PBE with that of LDA/PBEsol+$U$.
In Fig.~\ref{fig:lattice2}, we plot the deviations within PBE and PBEsol+$U$ from the experimental results.
It can be seen that PBEsol+$U$ performs much better than PBE.
For the complex $\alpha$-Np structure with seven lattice parameters, PBE could well reproduce the three lattice constants but cannot reproduce the four internal parameters. 
But all the seven lattice parameters could be reproduced by PBEsol+$U$.
The worst performance of PBEsol+$U$ lies in the $\gamma$-Pu structure, in which the largest deviation is 6.2\%.

\begin{figure}[t]
\begin{center}
\includegraphics[width=0.97\columnwidth]{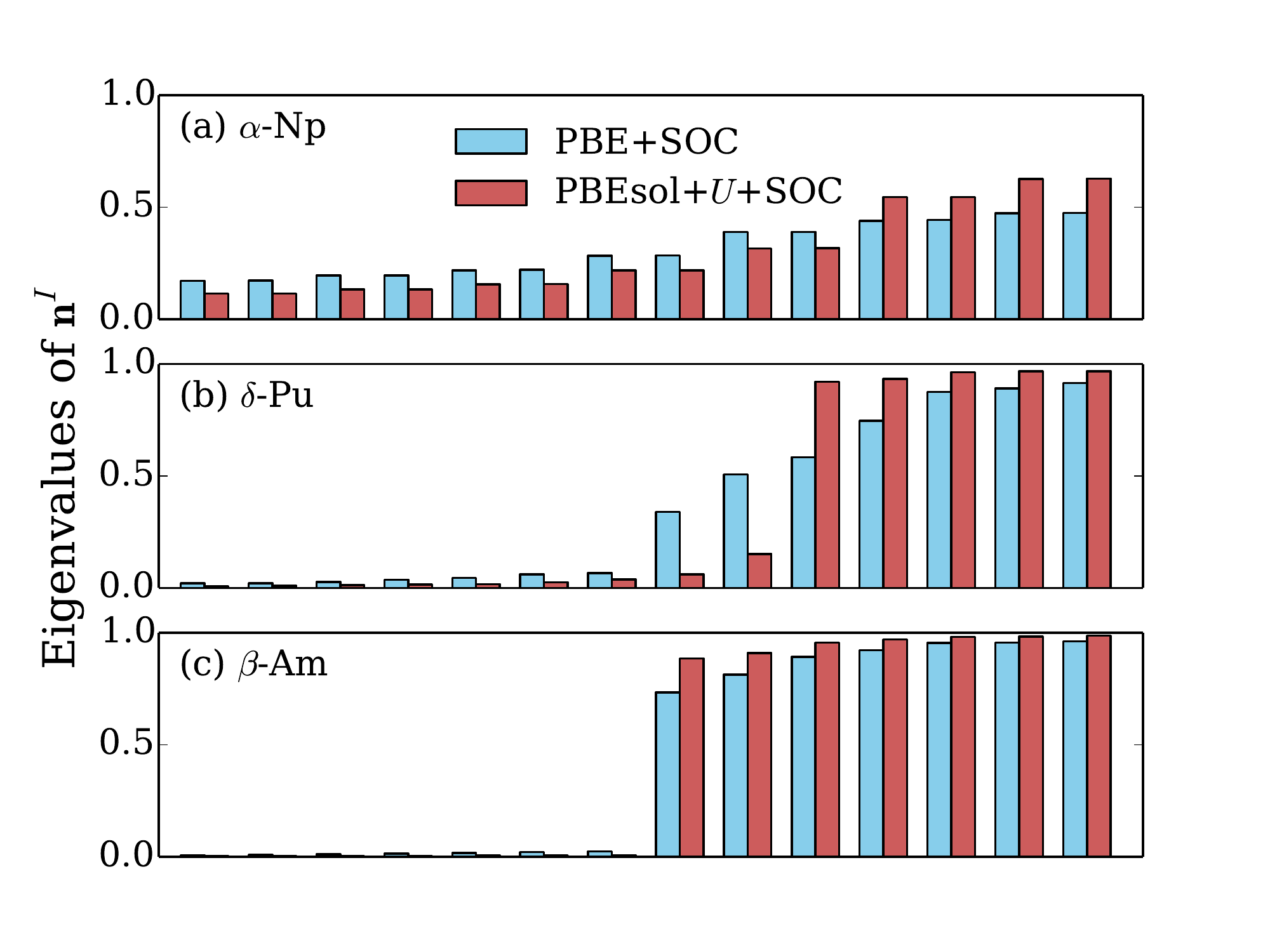}
\caption{The eigenvalues of the occupation number matrix (\ref{eq:occupation}) for (a) $\alpha$-Np, (b) $\delta$-Pu, and (c) $\beta$-Am.}
\label{fig:eigen}
\end{center}
\end{figure}

As concerns the bulk moduli $B$, the experimental results are limited and even some experimental data are controversial, such as that of $\alpha$-Pa~\cite{Blank2002}.
Thus here we only take $\alpha$-U and $\alpha$-Np as two examples.
For $\alpha$-U, our calculated $B$ within PBE+SOC (141.3 GPa) is in good agreement with that from previous literature~\cite{Soderlind2002,Richard2002,Bouchet2008,Taylor2008}.
But when comparing with the experimental values~\cite{Yoo1998} (135.5 GPa), our calculated $B$ within PBEsol+$U$+SOC (134.4 GPa) is closer than that within PBE+SOC.
For $\alpha$-Np, our calculated $B$ within PBE+SOC (192.6 GPa) and the previous theoretical values (see Ref.~\onlinecite{Richard2002} and reference therein) greatly overestimated the experimental value (118 GPa)~\cite{Dabos1987}, but our calculated $B$ within PBEsol+$U$+SOC (127.9 GPa) well reproduce the experimental result.
Therefore, we conclude that PBEsol+$U$+SOC with $U$ calculated from linear response approach is a reasonable choice for the description of the bulk properties of actinide metals.

From the calculated Kohn-Sham orbitals $\psi^\sigma_{{\bm k}\nu}$, it is straightforward to compute the occupation number matrix ${\bm n}^I$ (\ref{eq:occupation}) and the eigenvalues of ${\bm n}^I$ could be used to describe the degree of localization of 5$f$ electrons.
In Fig.~\ref{fig:eigen}, we plot these eigenvalues for $\alpha$-Np, $\delta$-Pu, and $\beta$-Am within PBE+SOC and PBEsolU+SOC.
The transition from itinerancy to localization of 5$f$ electrons along the actinide series is very clear.
Within PBEsol+$U$+SOC, all 5$f$ electrons of Np are itinerant while for Am, all the seven 5$f$ electrons of Am are nearly-localized.
The simple actinide physics is not clear within PBE+SOC.
The 5$f$ electrons of Pu are ``on the edge''~\cite{Lander2003}, giving rise to its complex physics.
From Fig.~\ref{fig:eigen} (b), about five 5$f$ electrons are localized, which yields a 5f$^5$ configuration for $\delta$-Pu.
This is in good agreement with the DMFT calculation~\cite{Savrasov2001,Shim2007},  X-ray absorption and photoemission experiments~\cite{Moore2009}. 

\section{Conclusion}
\label{sec:conclusion}

In the description of systems with strongly correlated and typically localized electrons, DFT+$U$ method is one of the widely used computational approaches to correct the inaccuracies of local density approximation (LDA), generalized gradient approximation of Perdew-Burke-Ernzerhof (PBE), and PBE revised for solids (PBEsol). 
For the actinide compounds with obviously localized 5$f$ electrons, such as dioxides, DFT+$U$ has been extensively used. 
But it is rarely used in the description of the actinide metals in which the 5$f$ electrons lie in between itinerancy and localization.
In addition, the important Coulomb parameter $U$ of actinide materials is usually determined empirically.
Thus particular emphasis was put on the necessity to compute the effective Coulomb interactions from first principles. 

The linear response approach to compute $U$ was described in detail and applied to the various allotropic phases of actinide metals under ambient pressure.
The effect of spin-orbit coupling, exchange-correlation functional and magnetic states are analyzed and found to be small.
The trend of $U$ along the series indicates the main feature of 5$f$ electrons, i.e., the transition from itinerancy to localization along the series.
The $U$ of $\alpha$-phase first decreases due to the decrease of atomic volume,
jumps from Pu to Am due to the change of the behavior of the 5$f$ electrons, 
and then increases monotonously for transplutonium metals.

In performing DFT and DFT+$U$ calculations on the bulk properties of actinide metals, PBEsol+$U$ could well reproduce a large amount of the lattice parameters and several bulk moduli and thus is the optimal choice.
In addition, the itinerant-localized 5$f$ electronic transition along the series could be clearly viewed from the occupation number matrix, which is defined as the projection of the Kohn-Sham orbitals into the localized orbitals set.
Therefore, we conclude that PBEsol+$U$+SOC with $U$ from linear response calculation is suitable for the description of actinide metals.

Nevertheless, DFT+$U$ approach has many inherent limits such as the static character and the calculated $U$ is not frequency-dependent.
But due to its internal self-consistency, DFT+$U$ with $U$ calculated from linear response approach represents a very useful computational tool to model actinide materials. 
It is able to significantly improve the conventional DFT and enable the possibility of calculations that would be extremely expensive for the quantum many-body method such as dynamical mean-field approximation.

\section*{Acknowledgments}

We would like to acknowledge the financial support from the  Science  Challenge  Project  of  China (Grant No. TZ2016004), National Science Foundation of China (Grant No. 21771167, 11874329, and 21601167), and the CAEP project (Grant No. TCGH0708).

\appendix
\section{Computational models}
\label{sec:model}

The linear response calculations of Hubbard $U$ are performed on the solid phases of actinide metals (Th-Cm, Cf) at ambient pressure.
Unless otherwise stated, the Hubbard $U$'s are calculated using the structures with the experimental lattice parameters, which are obtained from Pearson's handbook~\cite{Villars1997} except $\beta$-Pa~\cite{Marples1965} and $\beta$-Cf~\cite{Heathman2013}.

Thorium (Th) is the first element in the actinides series with empty 5$f$ orbitals for free atom but a substantial occupation of 5$f$ orbitals in its metallic condensed phase. 
Under ambient condition, Th has a close-packed face-centered cubic (fcc) structure ($\alpha$-Th) and at elevated temperature, Th transforms from fcc to body-centered cubic (bcc) at approximately 1673 K and bcc Th ($\beta$-Th) melts at approximately 2023 K~\cite{Chiotti1954}. The narrow 5$f$ bands, which could induce a Peierls distortion~\cite{Soderlind1995}, don't take effect at ambient pressure but play a role under high pressure. That is,  fcc Th transforms toward a body centered tetragonal (bct) structure at about 60 GPa~\cite{Vohra1991}. 

Protactinium (Pa) is the second element of actinides series with the electron configuration of free atom being [Rn]7$s^2$5$f^2$6$d^1$.
The 5$f$ electrons begin to play an important role and participate in the metallic bonding.
A low-symmetry bct structure is adopted by the solid-state phase of Pa under ambient condition ($\alpha$-Pa).
The melting point of Pa is about 1845 K and when approaching the melting point, there is another solid-state phase, $\beta$-Pa.
The crystal structure of $\beta$-Pa is controversial~\cite{Blank2002}.
Marples prepared Pa metal by reducing the tetrafluoride with calcium and predicted a bcc form of $\beta$-Pa from the extrapolation of thermal expansion data~\cite{Marples1965}.
Asprey {\it et al.} found a new fcc structure in a quenched arc-melted sample besides the bct structure of $\alpha$-Pu~\cite{Asprey1971}.
This fcc form of $\beta$-Pa was confirmed by the reversible transition between bct and fcc phases above 1473 K using both X-ray and impurity analyses~\cite{Bohet1978}.
For more details and discussion, one can refer to Ref.~\onlinecite{Blank2002}. 
Since the experimental lattice parameter of fcc Pa (5.018~\AA) is much larger than any theoretical values even within DFT+$U$, bcc is considered as the structure of $\beta$-Pa here.

\begin{figure}[b]
\begin{center}
\includegraphics[width=0.97\columnwidth]{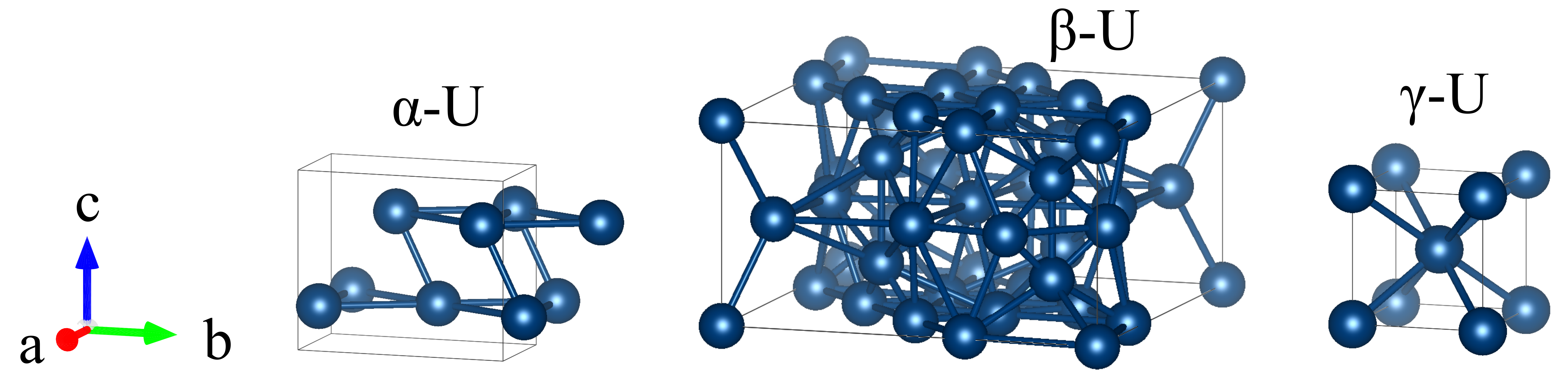}
\caption{Crystal structures of uranium.}
\label{fig:uranium}
\end{center}
\end{figure}

As the third element in the actinide series, uranium (U) has many very unique features that result from its 5$f$ electrons.
For example, the room-temperature orthorhombic crystal structure and the low-temperature charge density wave (CDW) transition of U are unique for an element at ambient pressure.
Neglecting the CDW phase, U exists in three solid-state phases at ambient pressure, which are labeled as $\alpha$, $\beta$, $\gamma$-U and shown in Fig.~\ref{fig:uranium}.
$\alpha$-U is orthorhombic with space group $Cmcm$ (No. 63) and all atoms are located at 4$c$ Wyckoff positions (0, $y$, 1/4). 
The structure is parameterized by the three lattice constants $a$, $b$, $c$, and the internal parameter $y$.
$\beta$-U is tetragonal with space group $P4_2/mnm$ (No. 136) and all atoms are located at five Wyckoff positions with seven internal parameters.
There are thirty atoms in the unit cell of $\beta$-U.
$\gamma$-U is bcc structure, which is adopted by most of metals when approaching the melting point for reasons of lattice stability.
At elevated temperatures, U transforms from $\alpha$ to $\beta$ at approximately 941 K and $\beta$ transforms to $\gamma$ at approximately 1048 K.

\begin{figure}[t]
\begin{center}
\includegraphics[width=0.97\columnwidth]{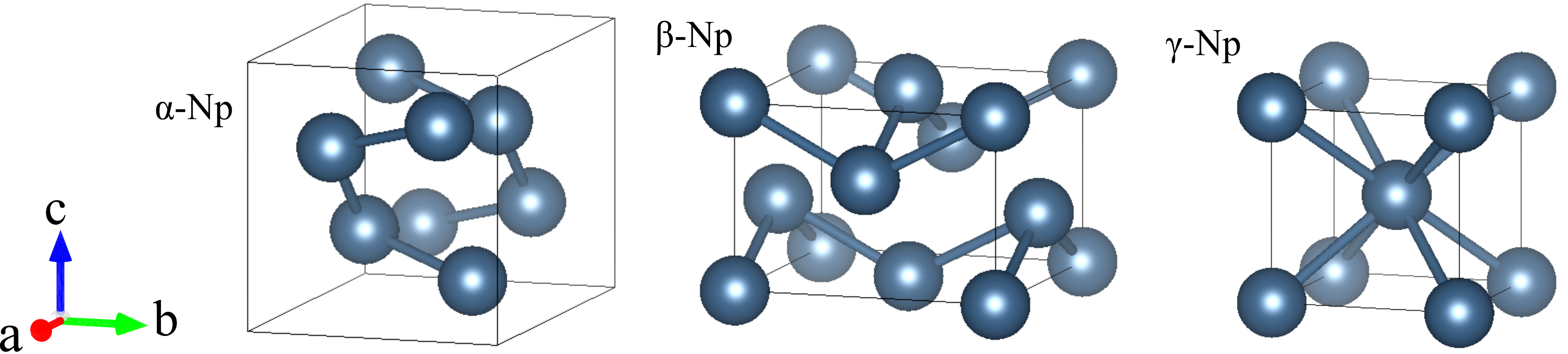}
\caption{Crystal structures of neptunium.}
\label{fig:neptunium}
\end{center}
\end{figure}

Neptunium (Np) is the fourth element in the series, which also exists in three solid-state phases at ambient pressure: 
orthorhombic $\alpha$, tetragonal $\beta$, bcc $\gamma$.
The structures are shown in Fig.~\ref{fig:neptunium}. 
The room-temperature $\alpha$-Np structure has space group $Pnma$ (No. 62), which is the subgroup of $Cmcm$.
All the Np atoms are also located at two 4$c$ Wyckoff positions ($x_{1,2}$, 1/4, $z_{1,2}$).
Thus a Peierls distortion along $a$-direction could make $\alpha$-U structure transition toward $\alpha$-Np structure, which has been theoretically demonstrated by our previous research~\cite{Qiu2016}.
The $\alpha$-Np structure is parameterized with the three lattice constants $a$, $b$, $c$, and four internal parameters.
The space group of $\beta$-Np is $P42_12$ (No. 90), which also has a lower symmetry than $\beta$-U.
But the structure of $\beta$-Np is much simpler.
The Np atoms are located at 2$a$ Wyckoff positions (0, 0, 0) and 2$c$ Wyckoff positions (0, 1/2, $z$). 
Only two lattice constants $a$, $c$ and one internal parameter $z$ are enough to characterize the $\beta$-Np structure.
At about 551 K, $\alpha$-Np transforms to $\beta$-Np and $\beta$-Np transforms to bcc Np ($\gamma$-Np) at about 823 K~\cite{Zachariasen1952}.

\begin{figure}[t]
\begin{center}
\includegraphics[width=0.97\columnwidth]{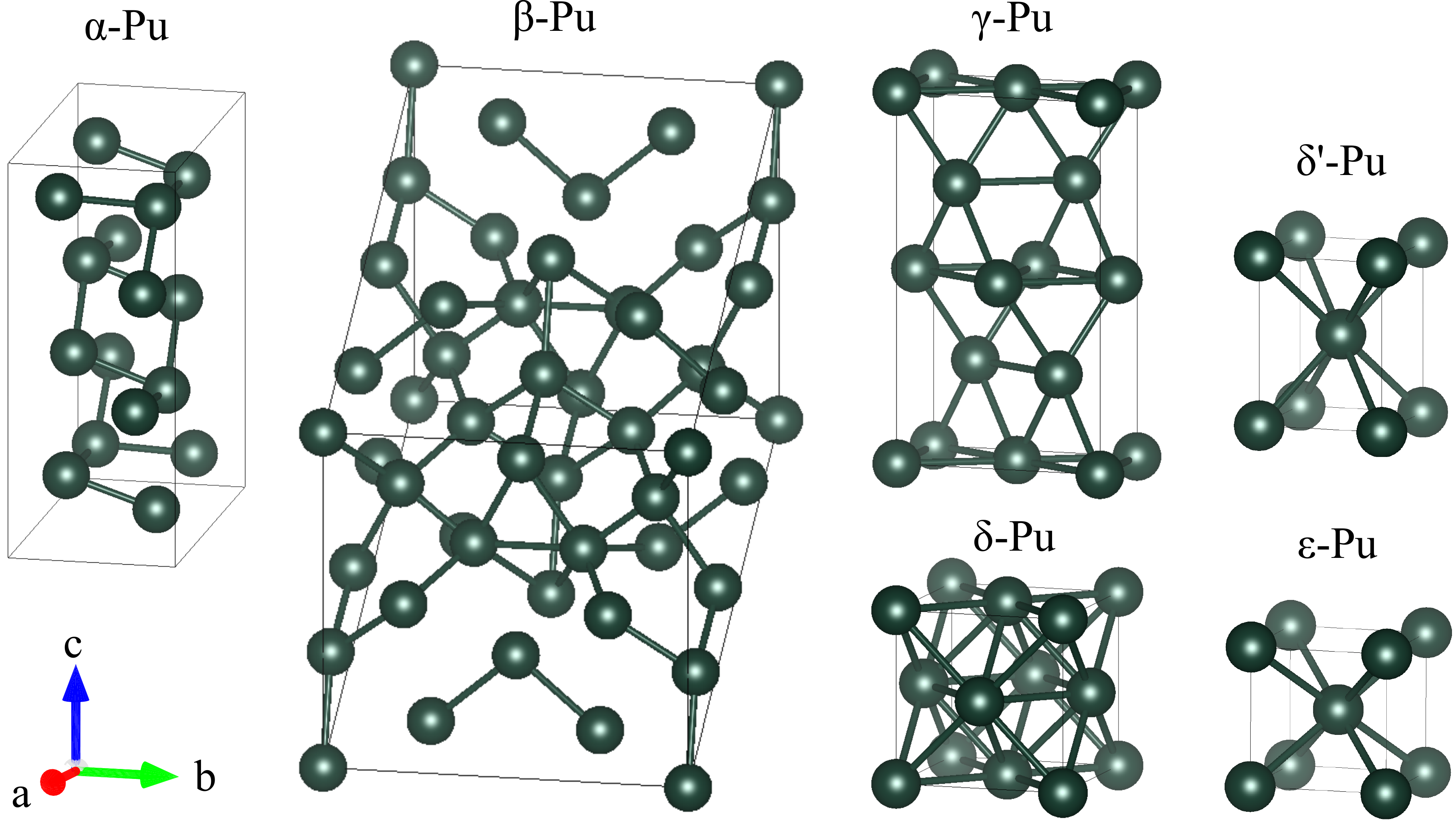}
\caption{Crystal structures of plutonium.}
\label{fig:plutonium}
\end{center}
\end{figure}

As the fifth element of actinides, plutonium (Pu) is the most complex element in the periodic table.
The electron configuration of Pu atom is [Rn]7$s^2$5$f^6$, which implies that only one electron is needed to reach a stable electron configuration, i.e., half-filled $f$-shell. 
On the other hand, there is spin-orbit splitting in the 5$f$ states and the $j=5/2$ states are filled for Pu atom.
Thus it has many states close to each other in energy but dramatically different in crystal structure, which is adopted in response to minor changes in its surroundings.
Before it melts at about 913 K, Pu undergoes six different phases, which is more than any other element.
See Fig.~\ref{fig:plutonium} for a sense of complexity.
$\alpha$-Pu is monoclinic with space group $P2_1/m$ (No. 11) and all atoms are located at eight 2$e$ Wyckoff positions ($x_{1\dots8}$, 1/4, $z_{1\ldots8}$).
$\alpha$-Pu transforms to $\beta$-Pu at about 395 K and $\beta$-Pu is also monoclinic with space group $C2/m$ (No. 12). 
$\beta$-Pu has thirty-four atoms per unit cell, i.e., the largest unit cell of pure metals.
$\beta$-Pu transforms to $\gamma$-Pu at about 479 K and $\gamma$-Pu is face-centered orthorhombic with space group $Fddd$ (No. 70).
Pu transforms from $\gamma$ phase to fcc $\delta$ phase at about 585 K and $\delta$-Pu has the lowest density.
$\delta$-Pu transforms to bct $\delta'$-Pu at about 724 K and $\delta'$-Pu transforms to bcc $\epsilon$-Pu at about 758 K.

For the transplutonium elements, i.e., americium (Am), curium (Cm), berkelium (Bk), and californium (Cf), all the metals are known to have a double hexagonal closed-packed (dhcp) $\alpha$ form and a high temperature fcc $\beta$ form at normal pressure.
There is a possible hexagonal closed-packed (hcp) phase of Cf, which is controversial~\cite{Heathman2013} and not considered here.
Research interests of transplutonium metals are focused on their high-pressure behavior since the pressure could induce the transition from localization to delocalization of 5$f$ electrons.
As can be seen from Fig.~\ref{fig:volume}, the atomic volume of $\alpha$-phase is nearly equal to or even greater than that of $\beta$-phase. 
This behavior results from their localized 5$f$ electrons at normal pressure.

\bibliographystyle{apsrev4-1}
\bibliography{an}
\end{document}